\documentclass[%
 reprint,
 amsmath,amssymb,
 aps,
]{revtex4-2}
\usepackage{graphicx}
\usepackage{dcolumn}
\usepackage{bm}
\usepackage{subcaption}
\usepackage{hyperref}

\usepackage{natbib}
\usepackage{ragged2e}
\usepackage{xcolor}
\begin{document}
\title{Vacuum Polarization Effects in Baryon-Loaded Magnetar Bursts
and Implications for X-ray Polarization}
\author{Tomoki Wada}
 \email{tomoki.wada@astr.tohoku.ac.jp}
\affiliation{Frontier Research Institute for Interdisciplinary Sciences, Tohoku University, Sendai, Japan}
\affiliation{Astronomical Institute, Graduate School of Science, Tohoku University, Sendai, Japan}
\affiliation{Department of Physics, National Chung Hsing University, Taichung, Taiwan}
\date{\today}
\begin{abstract}
Magnetars provide natural laboratories for strong-field quantum electrodynamics processes, such as vacuum polarization, which gives rise to vacuum resonance together with the plasma response.
We develop a general framework to describe vacuum resonance in a three-component plasma consisting of ions, electrons, and positrons, as expected in baryon-loaded magnetar bursts. 
By introducing a parametrization of the plasma composition, we establish the general criterion for the occurrence of vacuum resonance in such plasmas. 
Our analysis encompasses both Mikheyev-Smirnov-Wolfenstein-like adiabatic mode conversion and nonadiabatic eigenmode transition, highlighting their dependence on the plasma composition.
Applying this framework to baryon-loaded fireballs in magnetar bursts, we estimate the characteristic X-ray polarization signatures. 
Detection of these polarizations will provide observational signatures of vacuum polarization as well as baryon loading in magnetar fireballs.
\end{abstract}
\maketitle
\section{Introduction\label{sec:intro}}
%
Magnetars are neutron stars with magnetic fields that may
exceed the critical quantum electrodynamic (QED) field strength of 
$B_Q = m_e^2c^3/(e\hbar) = 4.4\times 10^{13}\,{\rm G}$,
where $m_e$ is the electron mass, $c$ is the speed of light,
$e$ is the elementary charge, and $\hbar$ is the reduced
Planck constant. Their persistent surface X-ray emission and bursting 
activity are interpreted as being powered by the intense magnetic 
fields~\cite{DunTho1992,HeyHer2005,HarLai2006,KasBel2017,EnoKis2019_neutronstar}.
A sudden release of magnetic energy near a magnetar surface can 
trigger the formation of a fireball, 
a plasma, consisting mainly of electron–positron pairs,
strongly coupled with photons
\cite{ThoDun1995,ThoDun1996}. These fireballs are capable of being
the source of bursting X-ray emission from magnetars.
Bursts are classified according to 
their total luminosity: $\lesssim 10^{44}\,{\rm erg\,s^{-1}}$ for 
short bursts (SBs), and $\gtrsim 10^{44}\,{\rm erg\,s^{-1}}$ for 
giant flares (GFs) \cite{KasBel2017}.
In addition to X-ray bursts, magnetars
have been proposed as sources of fast radio bursts (FRBs),
MeV gamma rays, cosmic rays, or ultrahigh-energy neutrinos
\cite[e.g.,][]{WadIok2021,WadShi2024,ShiWad2025,WadKim2025}.
Owing to their strong magnetic field, magnetars serve as natural 
laboratories for phenomena under strong magnetic fields \cite{Meszaros1992},
such as photon splitting \cite{HeyHer1997},
single-photon pair production, and
vacuum birefringence due to vacuum polarization \cite{TsaErb1975},
through their observations.

%
In strong magnetic fields, vacuum polarization 
predicted by non-linear QED theory \cite{HeiEul1936,Sch1951} 
modifies photon propagation \cite{GnePav1974,MesVen1979,HoLai2001}, particularly 
near the vacuum resonance, at which its influences become comparable to those of the plasma
\cite{GnePav1978,PavShi1979,SofVen1983,LaiHo2002}.
In the magnetar environment, both strong magnetic fields and 
plasmas are present \cite{GolJul1969}. While each 
contribution alone alters X-ray propagation, their 
combination can also significantly affect photon propagation through so-called
vacuum resonance. The vacuum resonance arises when the contributions from strong magnetic fields and plasmas become comparable. This phenomenon has been 
extensively studied in the context of magnetar atmospheric emission \cite{LaiHo2002,LaiHo2003,LaiHo2003_PRL,HoLai2003,HoLai2003_III,Oze2003,HoLai2004,NieBul2006,AdeLai2006,FerDav2011,KelZan2024},
where an electron-ion plasma is assumed to exist near the magnetar surface.
The vacuum resonance is also known to occur in relativistically streaming
electron-positron plasmas \cite{WanLai2007}.
In these previous works, the plasma was assumed to be either an electron-ion plasma or an electron–positron plasma, that is, to be a two-component plasma.

%
Recently, X-ray polarization measurements with the Imaging X-ray Polarimetry Explorer
(IXPE) \cite{IXPE2022} have opened a new window for probing magnetospheric and surface emission processes 
in magnetars. Energy-dependent X-ray polarization of magnetar persistent surface emission has been
observed in some magnetars 
\cite{TavTur2022,ZanTav2023,HeyTav2024,TurTav2023,RigTav2025,SteYou2025}. 
Remarkably, one of them, 4U 0142+6, exhibits about a 90-degree swing in the polarization 
angle in low energy and high energy \cite{TavTur2022}, a feature expected from vacuum 
resonance \cite{Lai2023}, although
resonant Compton scattering in a twisted magnetosphere can also produce 
such swings \cite{ThoLyu2002,TavTur2020}.
These X-ray polarization measurements of magnetars indicate that X-ray polarization from 
magnetars may serve as direct evidence of strong-field physics, beyond the 
critical QED field $B_Q$. Although polarization during the magnetar burst phase remains unexplored, 
it might be detected in the future by IXPE or planned 
eXTP \cite{GeJi2025}.

%
In magnetar bursts, their super-Eddington X-ray luminosity can 
drive baryon-loaded fireballs, thereby introducing
ion contributions to the dielectric properties of the plasma.
In most SBs, the fireball is confined by large-scale closed 
magnetic field lines, since its thermal energy 
density is much smaller than the magnetic energy density. 
At the footpoints of the trapped fireball, strong radiation pressure can
ablate baryons from the magnetar surface and load them into 
the fireball \cite{ThoDun1995,DemLyu2023}.
In contrast, during GFs, the thermal energy density of the fireball 
can exceed the magnetic field energy density, allowing it to expand
with little magnetic confinement, similar to fireball models for 
gamma-ray bursts \cite{Goo1986,Pac1986}. In this case, baryons can be 
loaded into the fireball, enhancing the kinetic energy of the associated 
plasma outflow \cite{ShePir1990,MesLag1993,MesRee2000}, 
as suggested in the observation of the GF from SGR~1806-20 
\cite{HurBog2005,MerGot2005,BogZog2007,FreGol2007}
and its radio afterglow \citep{GaeKou2005,CamCha2005,NakPir2005}.
In some of SBs, fireballs can expand along 
magnetic field lines if it forms near the magnetic pole
\cite{ThoDun2001}. The Galactic FRB associated with an SB from
the Galactic magnetar, SGR~1935+2154
\cite{Boc2020,CHIME2020_200428,Mer2020,Li2021,Rid2021,Tav2021}
may originate from such a polar fireball 
\cite{Iok2020,YanZha2021,WadAsa2025,LuKum2020,Kat2020,YamKas2022,WadIok2023}.
The large kinetic energy of a baryon-loaded fireball offers a 
favorable energy budget for this Galactic FRB \cite{WadIok2023}.
These theoretical expectations and observations 
both indicate the presence of baryons in the fireballs
in magnetar bursts; quantifying the baryon content is 
fundamental for understanding these bursting activities.

%
In this work, we derive the general condition for vacuum resonance 
in such baryon-loaded fireballs, and evaluate its effect
on the emergent X-ray polarization. 
The baryon loading in fireballs introduces a vacuum resonance, 
which is well studied in magnetar persistent surface emission 
\cite{LaiHo2002,LaiHo2003,LaiHo2003_PRL,HoLai2003,HoLai2003_III,Oze2003,HoLai2004,NieBul2006,AdeLai2006},
but not previously considered in detail for magnetar bursts.
This resonance occurs when the plasma and vacuum polarization
contributions to the dielectric tensor are comparable, enabling 
X-ray photons to convert between polarization eigenmodes and
thereby modifying the observed polarization.

%
This paper is organized as follows.
We introduce our model in Sec.~\ref{sec:model}. Sec.~\ref{sec:normal} 
presents an analysis of the photon normal modes.
In Sec.~\ref{sec:conv}, 
we investigate the propagation modes 
of photons over the vacuum resonance,
 Mikheyev–Smirnov–Wolfenstein (MSW)-like conversion 
(Sec.~\ref{sec:MSW}) and nonadiabatic mode transition
(Sec.~\ref{sec:nonad}) .
The observational expectations of these results are presented
in Sec.~\ref{sec:obs}.
Sec.~\ref{sec:sum} is devoted to the summary and discussion.
For simplicity, we do not consider the detailed baryonic composition
and treat all baryons as protons in this study.

\section{Normal Modes of Electromagnetic Waves in Baryon-Loaded Fireballs \label{sec:modes}}
%
We first describe the formation of a three-component 
(electron-positron-baryon) plasma, and photon propagation
in it, including the effect of vacuum polarization. 
The scenarios for baryon loading in magnetar bursts
and the setup of our model are described in Sec.~\ref{sec:model}.
We examine the normal modes of electromagnetic waves
in Sec.~\ref{sec:normal}.

\subsection{Baryon-Loading onto Fireballs in Magnetar Bursts \label{sec:model}}
%
We first describe the physical setup of our study: baryon-loaded 
fireballs in magnetar bursts. A sudden release of magnetic energy 
near the magnetar surface triggers the formation of a fireball.
Initially, the plasma may consist solely of electron-positron pairs created via
pair creation. However, baryons on the magnetar surface are ablated
by the photons and entrained into its plasma because
the luminosity exceeds the Eddington limit 
\cite{ThoDun1995,DemLyu2023}. Consequently, a baryon-loaded fireball 
naturally forms in a magnetar burst, and its plasma is described as
a three-component plasma.

\begin{figure}[tb]
\centering
\includegraphics[width=0.48\textwidth]{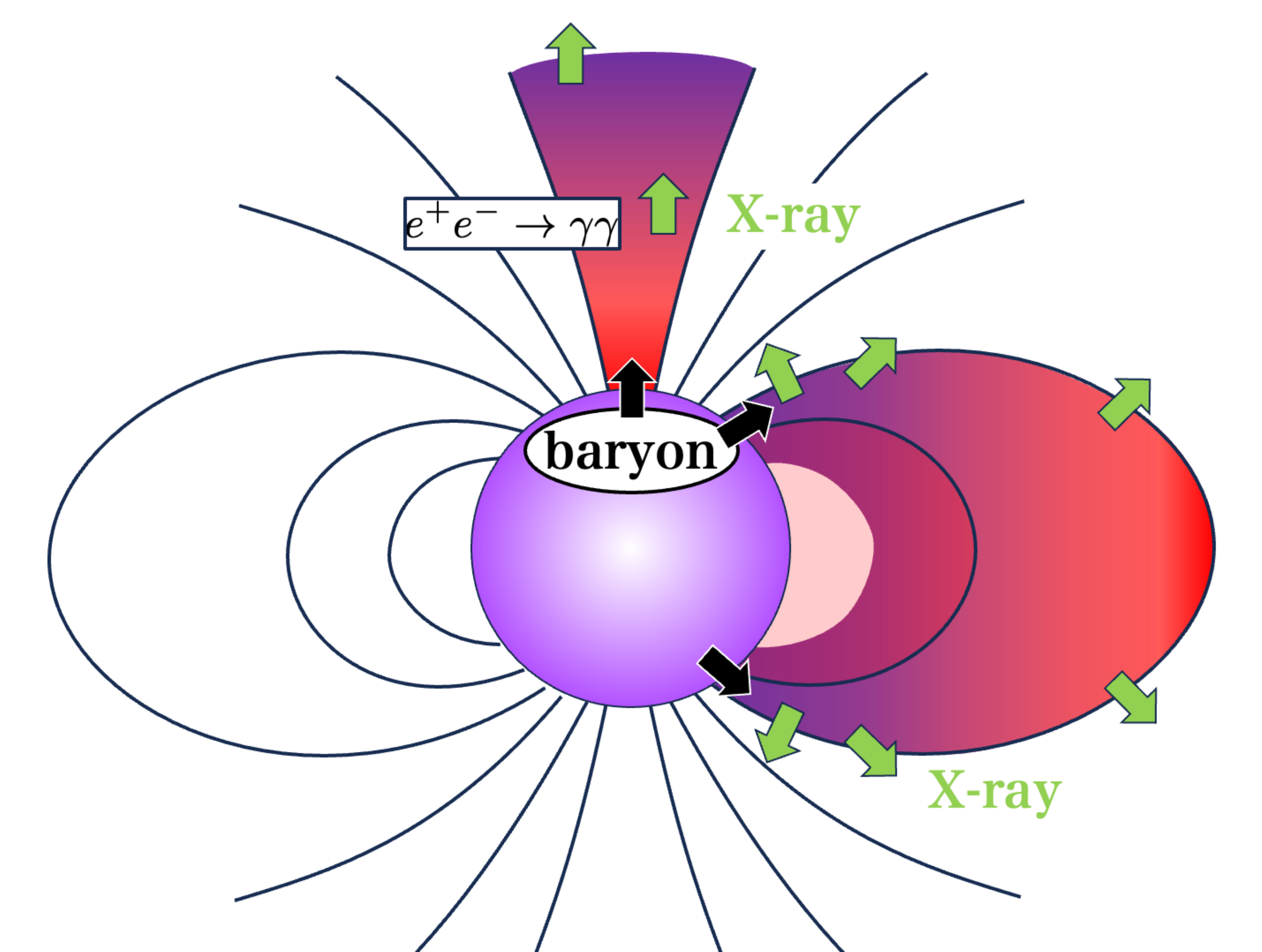}
\captionsetup{justification=raggedright,singlelinecheck=false}
\caption{\RaggedRight
Schematic picture of baryon-loaded fireballs in magnetar bursts;
a trapped fireball (right) and an expanding fireball (top). 
The coexistence of the two fireballs is not required.}
\label{fig:schematic}
\end{figure}

The composition of the three-component plasma depends on the geometry 
and location of the fireball (Fig.~\ref{fig:schematic}). 
In a trapped
fireball, the baryonic component 
would be distributed so as to maintain hydrostatic equilibrium 
between the radiation pressure and the magnetar's gravity, leading to 
a higher baryon contribution to the particle number density at the 
base of the closed magnetic field lines \cite{ThoDun1995}. 
In an expanding fireball,
the baryon contribution becomes dominant in the outer region, where the 
pair number density decreases relative to the inner regions due to 
pair annihilation. In both cases, baryon loading naturally occurs, 
and a strongly magnetized three-component plasma, consisting of 
baryons, positrons, and electrons, is realized 
(see Sec.~\ref{sec:thermal} for more details).

%
In order to characterize this three-component plasma, we define the
relevant variables. 
The number densities of electrons, positrons, and
baryons are denoted by $n_-$, $n_+$, and $n_b$, respectively. 
They satisfy the charge-neutrality condition
\begin{eqnarray}
n_-=n_++n_b.
\end{eqnarray}
We introduce a parameter $f$ to quantify the baryon content of the plasma,
\begin{eqnarray}
  f=\frac{n_b}{n_-},
  \label{eq:f}
\end{eqnarray}
such that the limit of $f\to 1$ corresponds to an electron-ion plasma, 
while $f\to 0$ corresponds to a purely electron-positron pair plasma.
The plasma frequency for each component is defined as
\begin{eqnarray}
\omega_{ps}&=&\sqrt{\frac{4\pi n_s e^2}{m_s}},
\end{eqnarray}
where $B$ is the strength of a magnetic field and 
$s = -,+,b$, and $m_s$ is the mass of species $s$. 
The corresponding cyclotron frequency is
\begin{eqnarray}
 \omega_{Bs}&=&\frac{eB}{m_sc}.
\end{eqnarray}
These frequencies satisfy the closure relation
\begin{eqnarray}
\omega_{p-}&=& (1-f)^{-1/2}\omega_{p+}
=(f\mu)^{-1/2}\omega_{pb}\\
\omega_{B-} &=& \omega_{B+}
=\mu^{-1} \omega_{Bb}
\end{eqnarray}
where $\mu=m_{\rm e}/m_{\rm p}\ll1$.

\subsection{Normal Modes of Electromagnetic Waves with Vacuum Polarization in a Three-Component Plasma \label{sec:normal}}
%
Dielectric tensor and inverse permeability tensor consist of
contributions from the standard plasma component and from the vacuum
polarization. 
Throughout this paper, the plasma is treated within the 
cold plasma approximation, whereby its thermal velocity is neglected.
In a coordinate system with the $z$-axis aligned with
the magnetic field, the dielectric tensor takes the form
\cite[e.g.,][]{Meszaros1992,LaiHo2002}
\begin{eqnarray}
  \bm{\varepsilon}=
 \begin{pmatrix}
  \epsilon& ig&0\\
  -ig &\epsilon&0\\
  0&0&\eta
 \end{pmatrix}
 +\begin{pmatrix}
   A_V-1& 0&0\\
  0&A_V-1&0\\
  0&0&A_V-1+Q_V
 \end{pmatrix},\nonumber\\
 \label{eq:varepsilon}
\end{eqnarray}
where the first term on the right-hand side corresponds to the 
plasma contribution of the three-component plasma, and the 
second term corresponds to the vacuum contribution. 
The quantities $\epsilon,g,\eta$ are given by 
\cite[e.g.,][]{Meszaros1992}
\begin{eqnarray}
  \epsilon&=&1+\frac{(2-f)v}{u-1}+\frac{f\mu v}{\mu^2u-1},
  \label{eq:epsilon}\\
  g&=&-\frac{fvu^{1/2}}{u-1}+\frac{\mu^2 f v u^{1/2}}{\mu^2 u-1},
  \label{eq:g}\\
    \eta&=&1-v(2-f+\mu f),
  \label{eq:eta}
\end{eqnarray}
where $u=\omega_{B-}^2/\omega^2$, and $v=\omega_{p-}^2/\omega^2$.
Throughout this paper,
we use the notation $u_s = \omega_{Bs}^2/\omega^2$ and 
$v_s = \omega_{ps}^2/\omega^2$ for $s = +,b$. 
The vacuum terms are \cite[e.g.,][]{HoLai2003}
\begin{eqnarray}
    A_V &\simeq& \begin{cases}
      1-2\delta_{\rm V}&(b\ll1)\\
    1+\frac{\alpha}{2\pi}\left[1.2-\frac{2}{3}\ln b-\frac{1}{b}(0.86+\ln b)-\frac{1}{2b^2}\right]
    &(b\gg 1)
  \end{cases},\nonumber\\
  \label{eq:a}\\
  Q_V &\simeq& \begin{cases}
  7\delta_{\rm V}&(b\ll1)\\
    -\frac{\alpha}{2\pi}\left[-\frac{2}{3}b+1.3-\frac{1}{b}(0.31+\ln b)-\frac{0.70}{2b^2}\right]    &(b\gg 1)    
  \end{cases},\nonumber\\
    \label{eq:q}
\end{eqnarray}
where $\delta_{\rm V} = \alpha b^2/(45\pi)$,
$\alpha$ is the fine-structure constant,
and $b = B/B_Q$.
The inverse permeability tensor is \cite[e.g.,][]{HoLai2003}
\begin{eqnarray}
  \bm{\mu}=
 \begin{pmatrix}
  1& 0&0\\
  0&1&0\\
  0&0&1
 \end{pmatrix}
 +
 \begin{pmatrix}
  A_V-1& 0&0\\
  0&A_V-1&0\\
  0&0&A_V-1+M_V
 \end{pmatrix},\nonumber\\
 \label{eq:mu}
\end{eqnarray}
where \cite[e.g.,][]{HoLai2003}
\begin{eqnarray}
M_V = \begin{cases}
-4\delta_{\rm V} & (b\ll1)\\
-\frac{\alpha}{2\pi}\left[\frac{2}{3}+\frac{1}{b}(0.15-\ln b)-\frac{1}{b^2}\right] & (b\gg1)
\end{cases}.
\end{eqnarray} In the weak-field limit, $b \to 0$, the 
vacuum terms satisfy $A_V-1\to0$, $Q_V\to0$, $M_V\to0$.
In this paper, we focus on waves with frequencies much larger
than the plasma frequency, that is, $v\ll 1$.
In the limit $f\to 0$, 
Eqs.~(\ref{eq:varepsilon})-(\ref{eq:eta}) reproduce to the dielectric 
tensor of pair plasma \citep[e.g.,][]{YanZha2015}, and in the limit
$f\to1$ and for $\mu\ll 1$, they reduce to the dielectric tensor of an
electron-ion plasma including vacuum contribution \cite[e.g.,][]{HarLai2006,HoLai2003}.
For simplicity,  as in \cite{LaiHo2002}, we neglect the wave damping, which
would introduce an imaginary contribution into the dielectric tensor
\cite{SofVen1983,Meszaros1992}.

%
The normal modes of propagating waves can be explicitly obtained 
as follows.  
For an electromagnetic wave $\vec{E}$ with frequency $\omega$ 
propagating along a newly defined $z$-axis, the propagation is
governed by the wave equation \cite[e.g.,][]{LaiHo2002,Meszaros1992},
\begin{eqnarray}
- k_\pm^2\vec{e}_z\times[\bm{\mu}\cdot(\vec{e}_z\times\vec{E}_\pm)]=\frac{\omega^2}{c^2}\bm{\varepsilon}\cdot\vec{E}_\pm,
  \label{eq:wave}
\end{eqnarray}
where $k_\pm$ is the wave vector and 
$\vec{e}_i$ ($i = x,y,z$) is a unit vector in the
direction of $i$. The subscripts ``$\pm$'' label the 
two eigenmodes for the quantities 
associated with the wave. 
Here, starting from the previous coordinate system,
where the tensor components are given by 
Eqs.~(\ref{eq:varepsilon})–(\ref{eq:eta}) and wave vector
is set to be on the $x$-$z$ plane, we perform a rotation in the $x$-$z$ 
plane such that the wave vector is aligned with the $z$-axis.
The unit eigenvector $\vec{E}_\pm$ can be expressed explicitly
in this rotated coordinate system, using the angle $\theta_{kB}$
between the magnetic field and the wave vector. We decompose
$\vec{E}_\pm$ as 
$\vec{E}_\pm = (\vec{E}_{\pm, {\rm T}}+E_{\pm, z}\vec{e}_z)/(1+E_{\pm,z}{}^2)^{1/2}$,
where $\vec{E}_{\pm, {\rm T}}$ is a normalized transverse component.
The transverse component is 
\begin{eqnarray}
    \vec{E}_{\pm, \rm T}=\frac{1}{(1+K_\pm^2)^{1/2}}\left(iK_\pm\vec{e}_x+\vec{e}_y\right),
 \label{eq:ET}
\end{eqnarray}
where these vectors are unit and mutually orthogonal.
By substituting these expressions and the rotated forms of
Eqs.~(\ref{eq:varepsilon}) and (\ref{eq:mu}) (rotated by $\theta_{kB}$
in the $x$-$z$ plane) into Eq.~(\ref{eq:wave}) and eliminating 
$E_{\pm, z}$, we obtain the eigenmodes of 
electromagnetic waves.
For the eigen modes, $K_\pm$ equals
\begin{eqnarray}
 K_\pm&=&\beta\pm\sqrt{\beta^2+R_V},
 \label{eq:Kpm}
\end{eqnarray}
where 
\begin{eqnarray}
  \beta=\frac{\left(-\epsilon_{\rm v}^2+\epsilon_{\rm v} \eta_{\rm v}(1+M_V/A_V)+g_{\rm v}^2\right)\sin^2\theta_{kB}}{2 g_{\rm v}\eta_{\rm v} \cos\theta_{kB}},
  \label{eq:beta_pre}
\end{eqnarray}
$\epsilon_{\rm v}= \epsilon+A_V-1$, $\eta_{\rm v}=\eta+A_V-1+Q_V$,
$g_{\rm v} = g$, and $R_V = 1+(M_V/A_V)\sin^2\theta_{kB}\simeq 1$.\footnote{
The sign of 
$g_V \eta_V\cos\theta_{kB}/(\epsilon_V\sin^2\theta_{kB}+\eta_V\cos\theta_{kB})$ 
in front of the square root is incorporated into the definition of $K_\pm$.}
The refractive indices of these eigenmode, $N_\pm = ck_\pm/\omega$, satisfy \cite[e.g.,][]{Meszaros1992}
\begin{widetext}
\begin{eqnarray}
  N_\pm^2 = \frac{R_V\epsilon_{\rm v}\eta_{\rm v}+
    \epsilon_{\rm v}^2 \sin^2\theta_{kB}+\epsilon_{\rm v}\eta_{\rm v}\cos^2\theta_{kB}-g_{\rm v}^2\sin^2\theta_{kB}
    \pm 2\eta_{\rm v}g_{\rm v}\cos\theta_{kB}\sqrt{\beta^2+R_V}}{2A_VR_V(\epsilon_{\rm v}\sin^2\theta_{kB}+\eta_{\rm v}\cos^2\theta_{kB})}.
  \label{eq:Npm}
\end{eqnarray}
\end{widetext}
Substituting Eqs.~(\ref{eq:epsilon})-(\ref{eq:eta}) into 
Eq.~(\ref{eq:beta_pre}) 
and using $A_V-1$, $\,Q_V,\,M_V,\,\mu\ll1$
give the explicit formula for $\beta$
in terms of wave frequencies as
\begin{widetext}
 \begin{eqnarray}
  \beta&=&\frac{u^{1/2}(1-u_b)\sin^2\theta_{kB}}{2\cos\theta_{kB}}\left(\frac{2-f}{f}\right)\left[1-\frac{(M_V+Q_V)(1-u^{-1})}{v(2-f)}\right].
    \label{eq:beta}
 \end{eqnarray}
 \end{widetext}
The factor of $(2-f)/f$ explicitly represents the effect of the three-component plasma.
Here, the divergence of $|\beta|$ as $f \to 0$ implies 
that the transverse components cannot be normalized according to 
Eq.~(\ref{eq:ET}): one of $x$ or $y$ components becomes zero.
As an example, for $\beta > 0$, 
at the limit of $f \to 0$ (i.e., $\beta \to \infty$), 
one finds $K_{+} \to \infty$ and $K_{-} \to 0$ (Eq.~\ref{eq:Kpm}), and hence 
$\vec{E}_{+} \parallel \vec{e}_{x}$ and $\vec{E}_{-} \parallel  \vec{e}_{y}$ (Eq.~\ref{eq:ET}).
This limit corresponds to the pair-plasma limit in which the eigenmodes of the wave 
separate into the O-mode, whose electric field lies in the $x$–$z$ plane, and the X-mode, 
whose electric field is perpendicular to that plane.
The factor $v(2-f) = v+v_+$ in the right-hand term represents the plasma frequency for the 
electron-positron pair component.
If an electromagnetic wave propagates adiabatically, i.e., remains in its 
eigenmode, the transverse components of the electric field are given by
Eqs.~(\ref{eq:ET}), (\ref{eq:Kpm}), and (\ref{eq:beta}).

The propagation eigenmodes are elliptically 
polarized and can be represented as the $+$ and $-$ modes, 
distinguished by the handedness of their polarization
(see Eq.~\ref{eq:Kpm}). In contrast, 
the electron response to the wave is primarily determined by the 
angle between the 
transverse component of the background magnetic field and the wave’s 
electric field. In this term, the modes are classified into X-mode 
($\vec{E} = \vec{e}_y$ in the basis of Eq.~\ref{eq:ET}) and
O-mode ($\vec{E} = \vec{e}_x$), 
both of which are eigenmodes in purely pair plasma.
Since the propagation eigenmodes do not coincide with the X/O-modes, 
photons can undergo conversion between the X- and O-modes
during adiabatic propagation.

\section{Adiabatic and Nonadiabatic Mode Transformations\label{sec:conv}}
When photons propagate in a three-component plasma, both adiabatic mode conversion 
and nonadiabatic mode transition can occur. 
In Sec.~\ref{sec:MSW}, we consider the adiabatic limit, in which a 
photon remains in the $+$/$-$ eigenmode, leading to 
conversion between X- and O-modes at the vacuum resonance. 
This phenomenon is formally analogous to the MSW
oscillation in neutrino physics. In Sec.~\ref{sec:nonad}, we 
investigate nonadiabatic transitions between $+$ and $-$
eigenmodes in the vicinity of the vacuum resonance.

\subsection{Adiabatic Mikheyev-Smirnov-Wolfenstein Mode Conversion around the Vacuum Resonance\label{sec:MSW}}
%
The vacuum resonance is defined as $\beta=0$ \cite{LaiHo2002},
that is, the transverse electric fields are circularly polarized,
$\vec{E}_{\pm, {\rm T}}\simeq (\pm i\vec{e}_x+\vec{e}_y)/\sqrt{2} $
(see Eq.~\ref{eq:ET}).
This condition is satisfied due to the comparable 
contribution from the vacuum polarization and the plasma at
(see Eq.~\ref{eq:beta}), 
\begin{eqnarray}
    (2-f)v_{\rm res} = (M_V+Q_V)(1-u^{-1}).
  \label{eq:vres}
\end{eqnarray}
For a given magnetic field, Eq.~(\ref{eq:vres}) contains
two independent variables, frequency $\omega$ and 
electron-positron number density $n_e = n_++n_- = (2-f)n_-$. 
For a fixed electron-positron number density, the condition in Eq.~(\ref{eq:vres}) gives the vacuum resonance frequency.
Conversely, for a photon with frequency $\omega$, this condition can be expressed in terms of the total pair number density
as
\begin{eqnarray}
    n_{e, {\rm res}} = \frac{m_e\omega^2}{4\pi e^2}(M_V+Q_V)(1-u^{-1}).
    \label{eq:nres}
\end{eqnarray}
 The wave propagating from an 
inner, high-density region to an outer, low-density region encounters the vacuum
resonance when Eq.~(\ref{eq:nres}) is satisfied. Around this density, 
MSW-like conversion occurs.
Stochastic nonadiabatic transitions between eigenmodes
will be  discussed in Sec.~\ref{sec:nonad}.
In the three-component plasma, the condition for vacuum resonance is determined 
solely by the total pair-particle number density, unlike 
the case of the magnetar surface emission 
where the condition is determined by the electron number density \cite{HoLai2001,LaiHo2002}.

\begin{figure}[tb]
\centering
\includegraphics[width=0.48\textwidth]{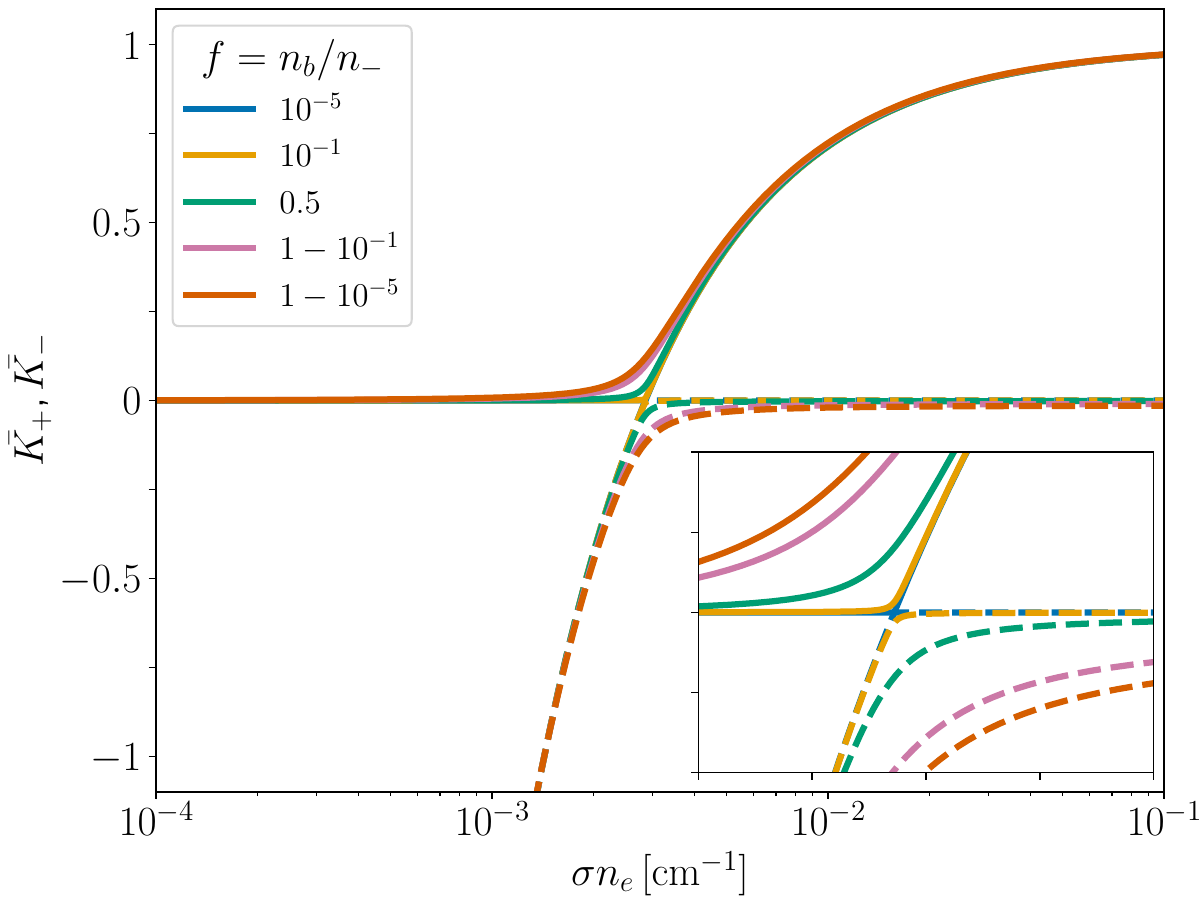}
\caption{\RaggedRight 
Electron-positron density dependence of $\bar{K}_\pm$, defined as $K_\pm$ 
normalized by the maximum
value of $K_+$ in the large $\sigma n_e$ limit. 
Solid lines correspond to $K_+$,
and dashed lines to $K_-$. Different colors indicate different values of 
$f$ (see Eq.~\ref{eq:f}). The curves for $f = 10^{-5}$, $10^{-1}$, $0.5$, 
$1-10^{-1}$, and $1-10^{-5}$ are further normalized by a factor of 
$1.6\times 10^6$, $1.6\times 10^2$, $2.5\times 10^1$, $1.0\times 10$, and $8.3$, 
respectively. The inset shows curves around the vacuum resonance in a linear scale.}
\label{fig:K}
\end{figure}

%
Figure~\ref{fig:K} shows the normalized $K_\pm$ for different values of $f$. 
The horizontal axis gives the electron-positron number density in unit of
$\sigma n_e$, where $\sigma = \sigma_{\rm T}\times\min\left(1,u^{-1}\right)$
\cite{CanLod1971,Her1979},
$\sigma_{\rm T}$ is the Thomson cross section, 
and cyclotron resonance is 
ignored for this normalization. 
The adopted parameters are $B = 10^{13}\,{\rm G}$, 
$\theta_{kB} = \pi/4$, and $\hbar\omega = 10\,{\rm keV}$.
The solid lines correspond to normalized $K_+$, and dashed lines represent 
normalized $K_-$. As the wave propagates from a high-density region to a 
low-density region, the direction of the $+$ mode electric field rotates from
the $x$ direction to the $y$ direction (see Eq.~\ref{eq:ET}), corresponding to
a transition from 
an O-mode to an X-mode. In contrast, the $-$ mode rotates from the $y$ direction
to the $x$ direction, that is, from an X-mode to an O-mode. 
As shown in Eq.~(\ref{eq:beta}), as $\beta$ decreases from a value $\beta \gg 1$, 
crosses zero, and continues decreasing to $\beta \ll -1$, the quantity 
$K_+$ ($K_-$) evolves from a large positive value, $2|\beta|$ (a small negative value, $-1/(2|\beta|)$), through $1$, and 
to a small positive value, $1/(2|\beta|)$ (a large negative value, $-2|\beta|$). 
The X/O-modes are conversed in this way.
This process is analogous to the MSW mechanism for neutrino 
oscillation: the photon remains in its $\pm$ eigenmode, and the X/O-modes 
are adiabatically converted. 

The behavior of $K_\pm$ varies with $f$.
For smaller $f$ values (nearly pair plasma), both elliptically polarized modes 
retain their handedness but approach linear polarization: the polarization 
ellipses become highly flattened, with the electric field predominantly along
the $x$ or $y$ axes.
In addition to the ellipticity, the gap between the $+$ and the $-$  modes 
is smaller than that of larger $f$ values.
This difference affects the probability of nonadiabatic mode transitions
at the vacuum resonance, which will be discussed in Sec.~\ref{sec:nonad}.

\subsection{Nonadiabatic Mode Transition around the Vacuum Resonance\label{sec:nonad}}
%
A photon propagating from a high-density region to a low-density region (from 
right to left in Fig.~\ref{fig:K}) can nonadiabatically 
transition from one eigen mode to the other.
In such transition, a photon in $\pm$ mode stochastically
jumps to $\mp$ mode, that is, from the solid (dashed) line 
to the dashed (solid) line in Fig.~\ref{fig:K}. Thus,
X(O)-mode in the high-density region remains
X(O)-mode in the low-density region across the vacuum resonance.
The conditions 
and probabilities for such nonadiabatic mode transitions are analyzed in this
subsection.

%
To quantify the adiabatic propagation and nonadiabatic mode transition, we 
compute the corresponding probabilities in a three-component plasma following
the procedure of \cite{LaiHo2002}. 
Rewriting $\vec{E}_{\pm,{\rm T}}$ in Eq.~(\ref{eq:ET}) as,
\begin{eqnarray}
\vec{E}_{+,{\rm T}} &
=& 
i\cos\theta_m\vec{e}_x+\sin\theta_m\vec{e}_y,\\
\vec{E}_{-,{\rm T}} &
=& 
-i\sin\theta_m\vec{e}_x+ \cos\theta_m\vec{e}_y,
\end{eqnarray}
a general polarized electromagnetic wave with frequency $\omega$ can be expanded
by these basis vectors as,
\begin{eqnarray}
  \vec{E}_{\rm T} = A_+(z)e^{i\phi_+}\vec{E}_{+,{\rm T}} + A_-(z)e^{i\phi_-}\vec{E}_{-,{\rm T}},
  \label{eq:Eexpand}
\end{eqnarray}
where $\phi_\pm = \int^z k_\pm dz$ and $A_\pm(z)$ is the amplitude function.
The mixing angle $\theta_m$ satisfies
\begin{eqnarray}
  \tan2\theta_m =\beta^{-1}.
  \label{eq:thetam}
\end{eqnarray}
We adopt the WKB approximation, in which $A_\pm$ and $\vec{E}_{\pm,{\rm T}}$ vary on a scale
much longer than the wavelength. Under this approximation, 
$ k_\pm \gg \left|{A_\pm'(z)}/{A_\pm(z)}\right|,
\left|E^j_{\pm,{\rm T}}{}'/E^j_{\pm,{\rm T}}\right|$~($j=x,y$ and $E^j_{\pm,{\rm T}}$ are the components of vector $\vec{E}_{\pm,{\rm T}}$), 
where a prime denotes $d/dz$.
Substituting Eq.~(\ref{eq:Eexpand}) into the wave equation (Eq.~\ref{eq:wave}), and 
applying the WKB approximation yields the following equation for $A_\pm$ 
\citep{LaiHo2002},
\begin{eqnarray}
 i\frac{d}{dz}
\begin{pmatrix}
{\mathcal A}_+\\{\mathcal A}_-
\end{pmatrix}
\simeq
\begin{pmatrix}
 -\frac{\Delta k}{2}&i\theta'_m\\
-i\theta_m'&\frac{\Delta k}{2}
\end{pmatrix}
\begin{pmatrix}
 {\mathcal A}_+\\{\mathcal A}_-
\end{pmatrix}.
\end{eqnarray}
Here, ${\mathcal A}_\pm = A_\pm(z) e^{i\phi_\pm}$, $\Delta k = k_+ - k_-$, and we have 
used $|\Delta k/k_\pm|\ll 1$, which follows from 
$v\ll1$ and $g\ll \epsilon,\eta$ 
(see Eqs.~\ref{eq:epsilon}-\ref{eq:eta}). A nonessential unity matrix has been subtracted.
As in \cite{LaiHo2002}, the modes evolve almost adiabatically if $|\Delta k/(2\theta_m')|\gg 1$
is satisfied. However, for
\begin{eqnarray}
  \gamma = \left|\frac{\Delta k}{2\theta_m'}\right|\ll1,
  \label{eq:gamma_pre}
\end{eqnarray}
which is easily satisfied for $\Delta k\sim0$ (see Eq.~\ref{eq:NpNm} later), 
the nonadiabatic transition becomes significant. 
This nonadiabatic conversion is a so-called Landau-Zener transition \cite{Lan1932,Zen1932}.

The probability for a nonadiabatic mode transition is determined by the adiabaticity 
parameter $\gamma$ at the vacuum resonance. Using
$\epsilon_{\rm v}, \eta_{\rm v} \gg g_{\rm v}$, which follows from $v\ll 1$, 
together with Eqs.~(\ref{eq:epsilon})-(\ref{eq:eta}), (\ref{eq:Npm}), and 
(\ref{eq:thetam}), we obtain the difference between the refractive indices for $+$ and 
$-$ modes as
\begin{eqnarray}
N_+-N_-&=&
-\frac{u^{1/2}}{u-1}\frac{fv}{1-u_b}\frac{\cos\theta_{kB}}{\sin2\theta_m}.
\label{eq:NpNm}
\end{eqnarray}
From Eqs.~(\ref{eq:thetam}), (\ref{eq:gamma_pre}), and (\ref{eq:NpNm}), $\gamma$ can be expressed in terms of
the relevant frequencies as
\begin{eqnarray}
 \gamma &=& 
 \frac{2\omega}{c}\frac{f^2}{\tan^2\theta_{kB}\sin^32\theta_m}\nonumber\\
 & & \times
 \frac{v^3}{uv'(1-u_b)^2(1-u^{-1})^2(M_V+Q_V)}.
 \nonumber\\
 \label{eq:gamma}
\end{eqnarray}
The probability of a nonadiabatic transition from one eigenmode to the other is then
given by the Landau-Zener formula \cite[e.g.,][]{Par1986,Hax1995}:
\begin{eqnarray}
    P_{\rm NA} = \exp\left(-\frac{\pi}{2}\gamma_{\rm res}\right),
    \label{eq:PNA}
\end{eqnarray}
where $\gamma_{\rm res}$ is the value of $\gamma$ evaluated at $v_{\rm res}$ 
(see Eq.~\ref{eq:vres}),
\begin{eqnarray}
 \gamma_{\rm res} = \frac{2\omega}{c}\frac{1}{\tan^2\theta_{kB}}
\left(\frac{f}{2-f}\right)^2
\frac{M_V+Q_V}{(1-u_b)^2}\frac{H_{\rm res}}{u},
\label{eq:gammares}
\end{eqnarray}
where $H_{\rm res} = v_{\rm res}/v'_{\rm res}$.
$\gamma_{\rm res}$ can be rewritten as $\gamma_{\rm res} = (2/\pi)(\omega/\omega_{\rm adi})^3$,\footnote{
The definition of $\omega_{\rm adi}$ is modified from that in \cite{LaiHo2002} by a factor of $(2/\pi)^{1/3}$,
so that $\omega = \omega_{\rm adi}$ corresponds to $P_{\rm NA} = \exp(-1)$. Exactly, Eq.~(\ref{eq:omegacri}) should be solved 
for $\omega$ appearing in $u_b$.}
where
\begin{eqnarray}
\frac{\hbar \omega_{\rm adi}}{m_ec^2}=\left[\frac{\hbar}{\pi m_e
c}\frac{\tan^2\theta_{kB}(1-u_b)^{2}}{H_{\rm res}(M_V+Q_V)}\left(\frac{B}{B_Q}\right)^2\left(\frac{2-f}{f}\right)^2\right]^{1/3}.\nonumber\\
\label{eq:omegacri}
\end{eqnarray}
The nonadiabatic transition becomes significant for photons with
$\omega<\omega_{\rm adi}$ 
The specific values of $\omega_{\rm adi}$ will be examined in Sec.~\ref{sec:obs}. 

%
The implications of nonadiabatic transitions in the extreme limits of $f$ are as follows.
In the limit $f \to 1$, the result reproduces that for the electron-ion plasma, as studied
in the context of persistent surface emission \cite{HoLai2003,LaiHo2002}. In this case, the nonadiabatic
transition probability depends explicitly on the photon energy. 
For intermediate $f$ values, as may 
occur in baryon-loaded fireballs in magnetar bursts, both of the adiabatic photon propagation 
and nonadiabatic mode transition can occur, resulting in mode changes from X(O)-mode to O(X)-mode
or preserving its X(O)-mode (see Sec.~\ref{sec:obs}).
In the opposite limit $f\to 0$ (purely pair plasma), $\gamma_{\rm res}$ converges to 0 (Eq.~\ref{eq:gammares}), 
yielding 
\begin{eqnarray}
\lim_{f\to 0}P_{\rm NA} = 1.
\label{eq:PNA1}
\end{eqnarray} 
Thus, an X(O)-mode photon retains its X(O)-mode, rather than its 
eigenmode, across the vacuum resonance. 
In this limit, MSW-like conversions do not occur, and 
the X- and O-eigenmode description in a pair plasma, where vacuum resonance is not expected,
is recovered due to nonadiabatic mode transitions.

\section{Observational Implication of Vacuum Resonance in Magnetar Bursts\label{sec:obs}}
%
Using the results in Sec.~\ref{sec:conv} and analytic models of baryon-loaded fireballs, we investigate the impacts of the vacuum 
resonance on the polarization properties of magnetar bursts. 
In Sec.~\ref{sec:thermal}, we show, from timescale considerations, that the thermal equilibrium plasma is realized in magnetar bursts.
The polarization of X-ray emission
from a trapped fireball is examined in Sec.~\ref{sec:trapped}, while that from an 
expanding fireball is considered in Sec.~\ref{sec:expanding}. 
The primary aim of this section is to demonstrate, within a simplified analytic framework, how vacuum resonance investigated in the previous sections influences the X-ray polarization from magnetar bursts. While detailed and global calculations are important, they are beyond the scope of this work.
The essential physics can 
be captured with the approach adopted here.
Throughout this section, we set $\hbar = 1$.

\subsection{Three-Component Plasma in Magnetar Bursts}\label{sec:thermal}
In magnetar bursts, which are intrinsically time-dependent phenomena, 
the fireball itself can exhibit variability.
Fireballs are thought to form through the dissipation of magnetohydrodynamic waves 
in the magnetosphere \cite{ThoDun1995}. A sudden displacement of the magnetar crust 
excites Alfv\'{e}n waves. For a crustal displacement of $\Delta \ell$, the 
characteristic frequency is estimated  as 
$\sim V_\mu/\Delta \ell \sim10^4\,{\rm s}^{-1}\,\Delta \ell_{,4}^{-1}\rho_{,11}^{1/6}$, 
where $\mu = 2\times10^{27}\,{\rm erg\,cm^{-3}} \rho_{,11}^{4/3}$ is the shear modulus of 
the crust \cite{ChaHae2008}, $\rho$ is the crustal density, $V_\mu =(\mu/\rho)^{1/2}$
is the sound speed in the crust, and the notation $Q_{,x} = Q/10^x$ denotes normalization
in cgs units. The shear modulus is evaluated assuming that the crust is isotropic
body-centered cubic polycrystal with atomic number 40 and mass number 120 \cite{ChaHae2008}.
$\Delta\ell \sim 10^{4}\,{\rm cm}$ is comparable to the hydrostatic pressure
scale height of the outer crust.
The fireball variability timescale is given by the inverse of this frequency, 
$t_{\rm var}\sim 10^{-4}\,{\rm s}\,\Delta \ell_{,4}\rho_{,11}^{-1/6}$. 
The injected Alfv\'{e}n waves dissipate through magnetospheric instability on 
the magnetospheric Alfv\'{e}n crossing timescale,
$t_{\rm diss}\sim r_{\rm fb}/c\sim 3\times 10^{-5}\,{\rm s}\,r_{\rm fb,6}$, 
where $r_{\rm fb}$ is the characteristic fireball radius and Alfv\'{e}n speed is 
assumed to be $c$. An optically thick electron-positron pair fireball thus forms
on this timescale, possibly 
modulated on $t_{\rm var}$.
Baryons are subsequently loaded from the magnetar surface with this variable timescale.

In trapped fireballs, the three-component plasma remains in hydrostatic and 
local thermal equilibrium. The hydrostatic equilibrium is established on a timescale of 
$t_{\rm hyd}\sim r_{\rm fb}/(c/\sqrt{3})\sim 2\times10^{-5}\,{\rm s}\,r_{\rm fb,6}$, 
where $c/\sqrt{3}$ is the sound speed in a radiation-dominated fluid. After baryons 
and their associated electrons are loaded into the pair fireball, the positrons will 
annihilate with these electrons. Near the photospheric radius, the pair-annihilation 
timescale is 
$t_{\rm ann}\sim 8/(3\sigma_{T}cn_-)\sim 1\times 10^{-7} \,{\rm s}\,T_{,10\,{\rm keV}}^2B_{,14}^{-2}r_{\rm fb,6}$,
where $T_{\rm ,10\,keV} = T/(10\,{\rm keV})$ is the temperature of the plasma.
Here, we have adopted the nonrelativistic
limit in which the product of cross section for pair annihilation and relative velocity 
is $3\sigma_{\rm T}c/8$.
For simplicity, we consider the photospheric radius where the optical depth for pair plasma is unity, i.e., 
$2n_-\sigma(T,B)r_{\rm fb}\sim 1$, where
$\sigma(T,B) =(4\pi^2/5)(T/(m_ec^2))^2(B/B_Q)^{-2}\sigma_{\rm T}$
is the Rosseland mean of the scattering cross section \cite{Meszaros1992,ThoDun1995}.
Because $t_{\rm ann}\ll t_{\rm hyd}< t_{\rm var}$, both hydrostatic and local thermal 
equilibrium would be maintained within the trapped fireball. 

In the simplest case of expanding fireballs, the three-component plasma remains in local thermal 
equilibrium, but hydrostatic equilibrium cannot be maintained due to its expansion 
at relativistic speeds. 
For expanding fireball, times and densities are evaluated in the plasma comoving frame. The dynamical timescale is 
$t_{\rm dyn} \sim r/(c\Gamma)\sim 1\times 10^{-5}\,{\rm s}\,r_{,6}\Gamma_{,0.4}$,
where $r$ is the radial distance from the magnetar center and $\Gamma$ is the bulk 
Lorentz factor. The pair-annihilation timescale is 
$t_{\rm ann}\sim 8/(3\sigma_{\rm T}cn_-)\sim 3\times 10^{-8} \,{\rm s}\,T_{,10\,{\rm keV}}^2B_{,14}^{-2}r_{\rm fb,6}\Gamma_{,0.4}^{-1}$,
where we have assumed $2n_-\sigma(T,B)r/\Gamma\sim 1$. The variability timescale in
the comoving frame is 
$t_{\rm var}\sim 3\times 10^{-5}\,{\rm s}\,\Delta \ell_{,4}\rho_{,11}^{-1/6}\Gamma_{0.4}^{-1}$.
Consequently, the condition $t_{\rm ann}\ll t_{\rm dyn}, t_{\rm var}$ is satisfied,
and local thermal equilibrium is achieved.

For fireballs expanding within a narrow magnetic flux tube, the timescale of photon 
diffusion perpendicular to the flux tube can be shorter than the dynamical timescale. 
In such cases, the fireball may evolve into a two-component structure: 
a baryon-loaded fireball confined to the initial flux tube and 
a surrounding pair-plasma fireball formed by the diffused photons \cite{WadIok2023}. 
In addition, temporal variations in the X-ray luminosity may 
lead to corresponding changes in the baryon fraction.
However, we do not consider these details in this work. 
Instead, we focus on cases where the baryon content is taken 
to be uniform in both space and time,
so that the effects of vacuum polarization can be clearly demonstrated.

\subsection{Polarization of Trapped Fireballs}\label{sec:trapped}
%
The X-ray
spectrum from trapped fireballs has been modeled in \cite{Lyu2002}, showing good 
agreement with observations. In a magnetized plasma, the scattering cross section for X-mode
photons is suppressed by a factor of $\sim (\omega/\omega_{B-})^2 = u^{-1}$ for $\omega<\omega_{B-}$,
resulting in a spectrum almost composed of X-mode photons \cite{Lyu2002}.
Consequently, lower-energy X-ray photons escape from the inner region of the fireball with a 
higher temperature, producing a spectrum flatter than a blackbody. \cite{Lyu2002} 
demonstrated that the emergent X-ray spectrum can be characterized by a single parameter,
the brightness temperature $T_b$.

%
In the context of the vacuum resonance, the relative location of the photospheric radius 
and the vacuum resonance determines the resulting polarization \cite{LaiHo2002}.
In a fireball with a given $T_b$, photons of energy $\omega$ escape from the depth where 
the temperature is \cite{Lyu2002}
\begin{eqnarray}
  T_\omega = T_b\sqrt{1+\frac{3\pi^2}{5}\left(\frac{T_b}{\omega}\right)^2}.
  \label{eq:Tw}
\end{eqnarray}
At the location, the baryon number density 
should be around the value obtained from the hydrostatic equilibrium condition \cite{ThoDun1995},
and is parametrized by ${\mathcal F}_{\rm trap}$ as
\begin{eqnarray}
  n_{b,{\rm trap}} = - \frac{{\mathcal F}_{\rm trap}}{m_bg_{\rm NS}}\frac{d}{dr} \left(\frac{a_{\rm rad}}{3}T_\omega^4\right).
  \label{eq:nbtrap}
\end{eqnarray}
Here, ${\mathcal F}_{\rm trap}$
is the baryon number density normalized by the value that can be supported 
by photon pressure against gravity,
$g_{\rm NS}$ is 
surface gravitational acceleration of magnetar,
and $a_{\rm rad} = \pi^2/(15c^3)$ \cite[e.g.,][]{RybickiLightman1979}.
${\mathcal F}_{\rm trap} = 1$ corresponds to the case of hydrostatic equilibrium.
Based on timescale considerations in Sec.~\ref{sec:thermal}, hydrostatic equilibrium is 
expected to be established. Nevertheless, we retain ${\mathcal F}_{\rm trap}$ as a free parameter
to allow for the possibility that this assumption can be tested 
through future X-ray polarization measurements.
We set the magnetar mass to be $1.4M_\odot$ and its radius to be $10\,{\rm km}$.
The electron and positron number densities $n_{\rm \pm,trap}$ are obtained from the thermal 
equilibrium condition, as \cite[e.g.][]{Landau1980stat,WadIok2023}
\begin{eqnarray}
  n_{\pm,{\rm trap}} &=& \mp \frac{n_{b,{\rm trap}}}{2}+
  \left[\frac{n_{b,{\rm trap}}^2}{4}\right.  \nonumber\\
    &+&\left. 4 \max\left(1,\frac{m_e^2c^4 b^2}{T_\omega^2}\right)
    \left(\frac{m_e T_\omega}{2\pi}\right)^3
    e^{-2m_ec^2/T_\omega}\right]^{1/2}.
    \label{eq:npmexp}\nonumber\\
\end{eqnarray}
For a fireball with a given ${\mathcal F}_{\rm trap}$, for each photon energy, we obtain the temperature at the photospheric radius using
Eq.~(\ref{eq:Tw}), and then obtain $f$ using Eqs.~(\ref{eq:f}), (\ref{eq:nbtrap}), and 
(\ref{eq:npmexp}).
Using $\omega$, the magnetic field, and Eqs.~(\ref{eq:nres}) and (\ref{eq:gammares}), 
we evaluate $n_{e,\,\rm res}$ and $\gamma_{\rm res}$ for the resonance condition.
In the following, $n_{+,\,{\rm trap}}+ n_{-,\,{\rm trap}}$ is denoted as $n_{e,\rm trap}$.
The critical photon energy $\omega_{\rm cri}$ is defined as the energy below which all photons
experience vacuum resonance, that is, $n_{e,{\rm trap}}>n_{e,{\rm res}}$ is satisfied. 
\begin{figure}[tb]
    \includegraphics[width=\linewidth]{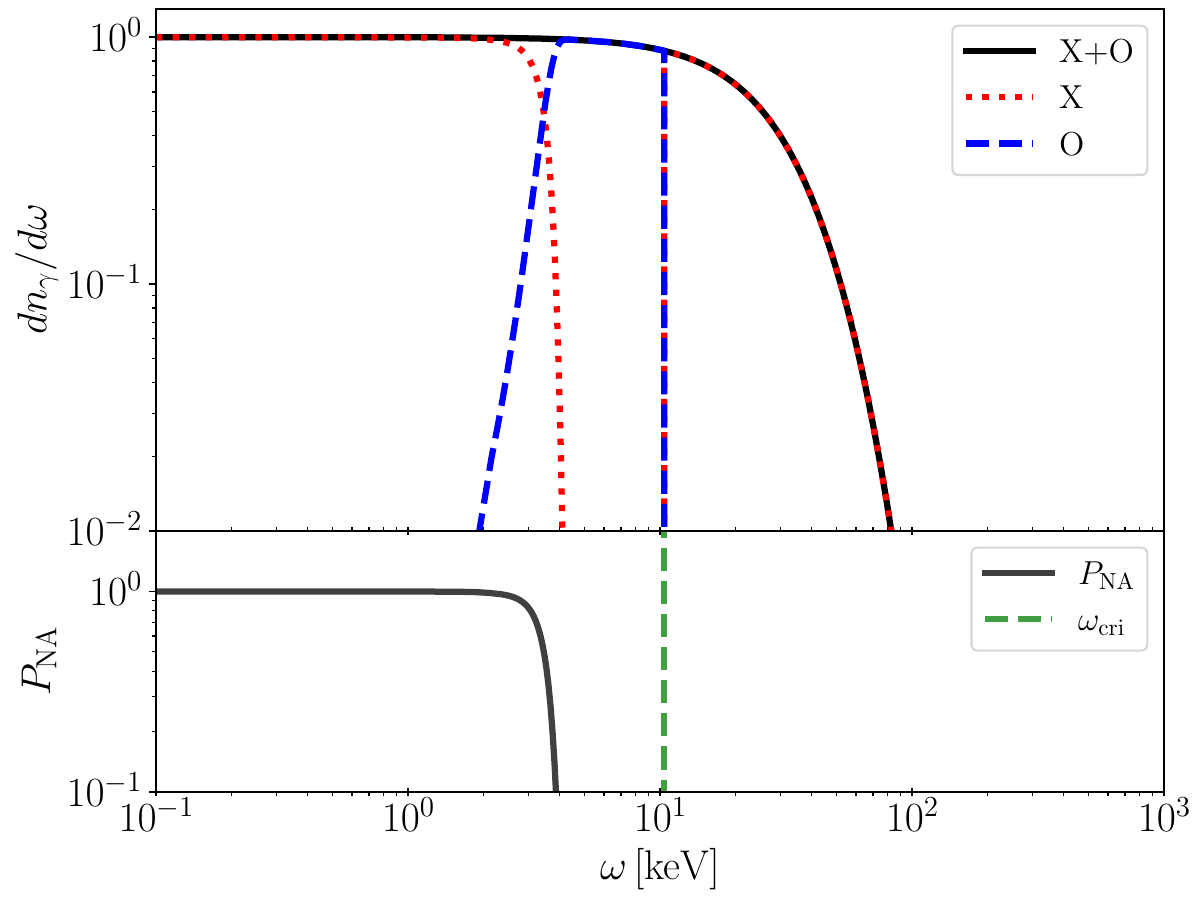}
    \includegraphics[width=1.07\linewidth,trim = 1.5cm 0cm 0cm 0cm ,clip]{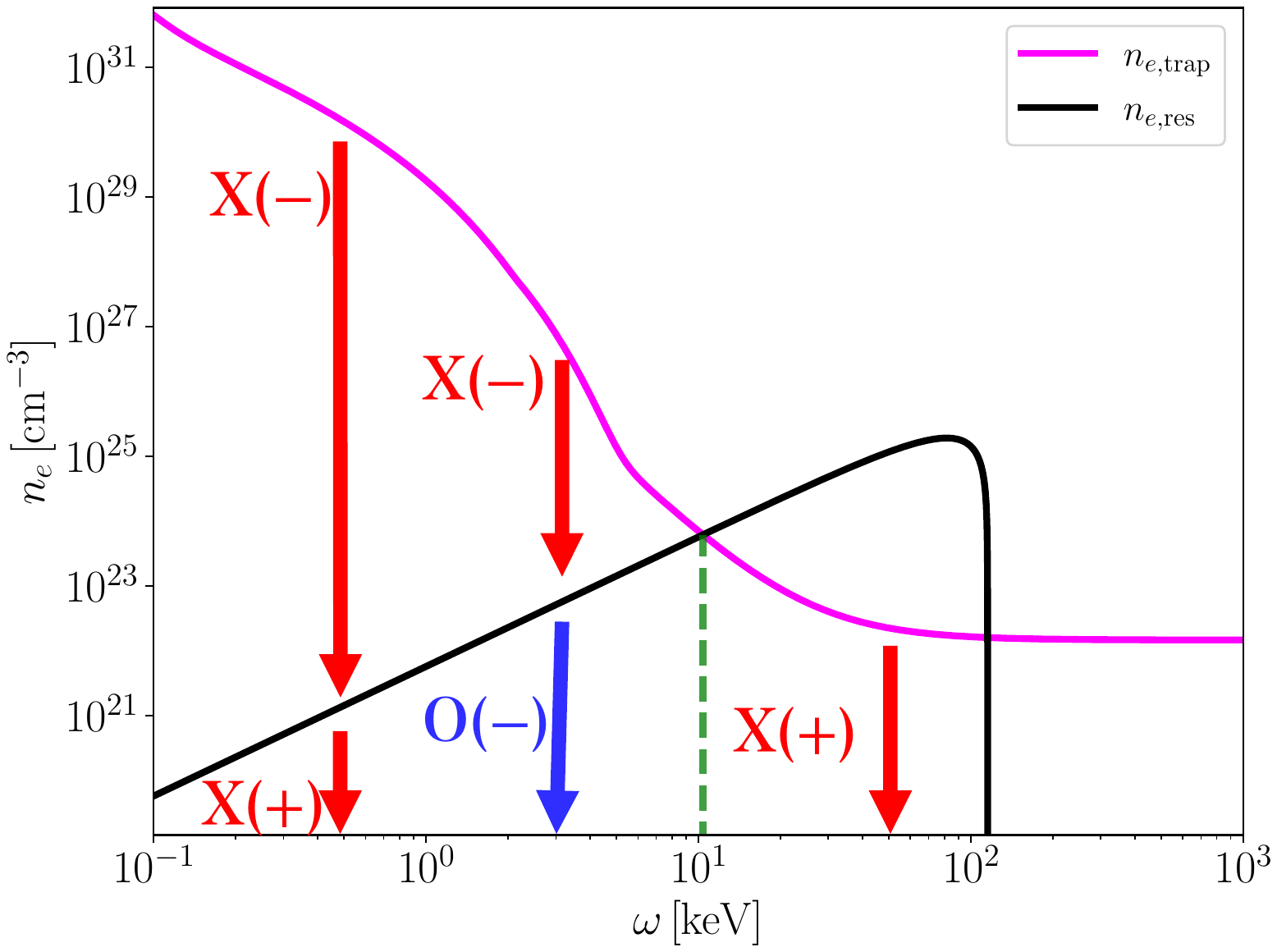}
\caption{\RaggedRight
Observed spectrum from a trapped fireball and pair number densities. 
(Top panel) The black line shows the total observed spectrum, the red dotted line indicates the 
observed X-mode photons, and the blue dashed line indicates the observed O-mode photons. 
The lower portion shows the nonadiabatic transition probability. The green vertical line marks
$\omega_{\rm cri}$. 
(Bottom panel) The magenta line shows the number density at which X-mode photons are emitted from 
the plasma, and the black line shows $n_{e,{\rm res}}$
where vacuum resonance occurs. The red and blue
arrows indicate the propagation of photons and their typical modes in terms of X/O, rather than 
$\pm$.
}
\label{fig:trapped}
\end{figure}

%
The bottom panel of Fig.~\ref{fig:trapped} shows the evaluated pair number density 
$n_{e,\rm trap}$ (magenta line) together with the resonance number density $n_{e,{\rm res}}$
(black line). The adopted parameters are $B=10^{13}\,{\rm G}$, $T_b = 10\,{\rm keV}$, 
${\mathcal F}_{\rm trap}= 10$ for fireball, $\theta_{kB}=\pi/4$, $H_{\rm res} = 10\,{\rm cm}$ 
for propagation. X-mode photons are emitted at $n_{e,\rm trap}$ and
propagate into a lower-density region \cite{Lyu2002}. 
The green vertical 
dashed line marks $\omega_{\rm cri}$.
Sudden decrease of the black line around $100\,{\rm keV}$ is because of cyclotron resonance,
i.e., $u = 1$ (see Eq.~\ref{eq:nres}).
The modes of propagating photons are also shown by red and blue arrows. 

The observed photon spectra in X- and O-mode should be evaluated, taking into account
the adiabatic and nonadiabatic mode transformations.
For $\omega\geq\omega_{\rm cri}$ except $\omega\simeq \omega_{B-}$, X-mode photons 
emitted as $-$ mode ($\omega\gtrsim\omega_{B-}$) or as $+$ mode ($\omega\lesssim\omega_{B-}$)
are observed as X-mode. The observed spectra are
\begin{eqnarray}
    \frac{dn_{\gamma, {\rm X}}}{d\omega} (\omega) &=& \frac{dn_{\rm BB}}{d\omega}(T_\omega)
    \label{eq:dndwlx}    \\
    \frac{dn_{\gamma, {\rm O}}}{d\omega} (\omega) &=& 0,
\end{eqnarray}
where $n_{\gamma,{\rm I}}$ (I$=$X/O) is the number density of photon in X/O-mode, and 
$dn_{\rm BB}/d\omega(T_\omega)$ is the photon spectrum of the blackbody radiation with 
temperature $T_\omega$ \cite{Lyu2002,RybickiLightman1979}.
The possible absorption and scattering are ignored in this work for simplicity.
For $\omega<\omega_{\rm cri}$, X-mode photons are emitted as the $-$ mode (see 
Fig.~\ref{fig:trapped}), and undergo a nonadiabatic transition with 
probability of $P_{\rm NA}$ in Eq.~(\ref{eq:PNA}).
Photons that undergo nonadiabatic transitions transform into the $+$ mode and are observed
as X-mode, while adiabatically propagating $-$ mode photons are observed as O-mode, 
due to the MSW-like conversion. 
Thus, the observed X- and O-mode photons are expressed as 
\begin{eqnarray}
    \frac{dn_{\gamma, X}}{d\omega} (\omega) &=& P_{\rm NA} \frac{dn_{\rm BB}}{d\omega}(T_\omega)\\
    \frac{dn_{\gamma, O}}{d\omega} (\omega) &=&  (1-P_{\rm NA}) \frac{dn_{\rm BB}}{d\omega}(T_\omega),
    \label{eq:dndwso}
\end{eqnarray}
where $P_{\rm NA}$ is given by Eq.~(\ref{eq:PNA}) for each $\omega$.

%
The top panel of Fig.~\ref{fig:trapped} shows the observed spectrum and polarization. Nonadiabatic 
transitions efficiently occur in the photon-energy range where both
$\omega<\omega_{\rm cri}$ and $P_{\rm NA}\sim1$ are satisfied.
In this energy regime, X-mode photons emitted from the inner region undergo a nonadiabatic 
transition from their $-$ mode to a $+$ mode, and are therefore observed as X-mode photons (see 
also Fig.~\ref{fig:K} and the bottom panel of Fig.~\ref{fig:trapped}). Adiabatic propagation 
takes place in the photon-energy range where
$P_{\rm NA}\sim 0$ is satisifed. In this range, 
for photons with $\omega<\omega_{\rm cri}$, the 
emitted X-mode photons propagate adiabatically and, through MSW-like mode conversion, 
escape as O-mode photons. 
The adiabatic propagation in this energy range can render the X-ray polarization distinguishable 
from other energy ranges 
(top panel of Fig.~\ref{fig:trapped}).

\begin{figure}[tb]
    \includegraphics[width=\linewidth]{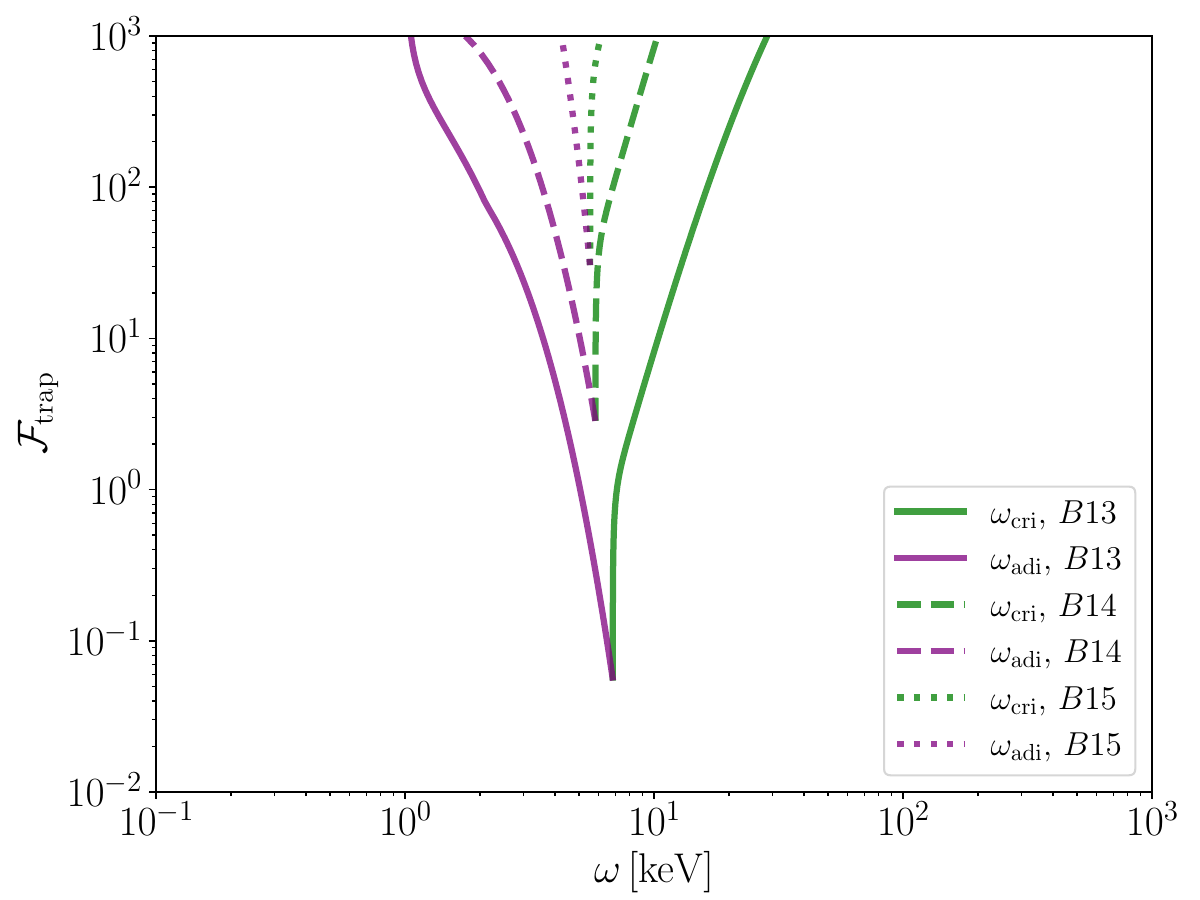}
\caption{\RaggedRight
$\omega_{\rm cri}$ (green line) and $\omega_{\rm adi}$ (purple line) for different baryon loading, ${\mathcal F}_{\rm trap}$. The legends $Bx$ represents the model with $B = 10^{x}\,{\rm G}$.}
\label{fig:trap_para}
\end{figure}

Figure~\ref{fig:trap_para} shows the ${\mathcal F}_{\rm trap}$ dependences of 
$\omega_{\rm cri}$ (green lines) and $\omega_{\rm adi}$ (purple lines).
The adopted parameters, except ${\mathcal F}_{\rm trap}$ and $B$, are the 
same as in Fig.~\ref{fig:trapped}. 
The solid, dashed, and dotted lines represent these energies for
$B = 10^{13}\,{\rm G}$, $10^{14}\,{\rm G}$, and $10^{15}\,{\rm G}$, respectively.
Values of ${\mathcal F}_{\rm trap}$ are taken from $10^{-2}$ to $10^3$, 
with the hydrostatic equilibrium satisfied at ${\mathcal F}_{\rm trap}\sim1$
(see Eq.~\ref{eq:nbtrap}). 
For a fireball with given ${\mathcal F}_{\rm trap}$, the MSW-like adiabatic 
mode conversion from X-mode to O-mode occurs in the energy range between the 
purple and green lines.
For the stronger magnetic fields,
the energy range of MSW-like conversion becomes narrower due to lower 
$\omega_{\rm cri}$ and higher $\omega_{\rm adi}$ 
(Eqs.~\ref{eq:a}, \ref{eq:q}, \ref{eq:omegacri}).
For a fireball with ${\mathcal F}_{\rm trap}$ smaller than the intersection of purple and green lines,
MSW-like conversion does not occur, and all photons are emitted in the X-mode. 
This is equivalent to the bayron-free pair fireball in the limit of $f\to 0$.
We note that, in such fireballs, all X-mode photons with 
$\omega<\omega_{\rm cri}$ undergo nonadiabatic mode transition with probability 
$P_{\rm NA} \simeq 1$ (see Eq.~\ref{eq:PNA1}). 
The modification to the polarization becomes more significant 
for fireballs with more baryons and weaker magnetic fields.

\subsection{Polarization of Expanding Fireballs along Open Magnetic Fields}\label{sec:expanding}
%
The dynamics of baryon-loaded expanding fireballs along open magnetic field lines have 
been modeled in \cite{WadIok2023}. An expanding fireball is characterized by the surface 
magnetic field, $B_0$, the initial
temperature $T_0$, the initial fireball size $l_0$, and the initial amount of baryon, 
described by ${\mathcal F}_{\rm exp}$.\footnote{
${\mathcal F}_{\rm exp}$ corresponds to $\eta^{-1}$ in \cite{WadIok2023}.} 
A dipole magnetic field is assumed here.
The total luminosity of the fireball is given by $\pi l_0^2ca_{\rm rad}T_0^4$ \cite{Iok2020}, 
and the initial rest-mass energy density of baryons is set to be
${\mathcal F}_{\rm exp} a_{\rm rad}T_0^4$.

%
In this work, we focus on the radiation-dominated fireball.
For the fireballs, the radial 
dependence of comoving temperature, Lorentz factor, and comoving baryon density 
are \cite{WadIok2023}
\begin{eqnarray}
    T &=& T_0\bar{r}^{-3/2}\label{eq:Texp},\\
    \Gamma &=& \bar{r}^{3/2},\\    
    n_{b, {\rm exp}}&=&  {\mathcal F}_{\rm exp}\frac{a_{\rm rad}T_0^4}{m_bc^2} \bar{r}^{-9/2},
    \label{eq:rhoexp}
\end{eqnarray}
where $\bar{r}$ is the radial coordinate normalized to the magnetar radius. 
The comoving pair number density is obtained from Eq.~(\ref{eq:npmexp}) with the 
subscript ``trap" replaced by ``exp," together with Eqs.~(\ref{eq:Texp}) and 
(\ref{eq:rhoexp}).  X-mode photons with energy $\omega$ are emitted at the radius
where
\begin{eqnarray}
    n_{e,{\rm exp}} \sigma_{\rm T}\min\left(1,u^{-2}\right)\frac{r}{\Gamma} = 1,
    \label{eq:tau_exp}
\end{eqnarray}
is satisfied \cite{AbrNov1991}.
For each $\omega$, by solving Eq.~(\ref{eq:tau_exp}) for $\bar{r}$, and using Eq.~(\ref{eq:Texp}), 
we obtain the plasma temperature $T_\omega$ at the photospheric radius.
The emitted photon spectrum at $\omega$ is obtained from a blackbody radiation with 
temperature $T_\omega$. Using Eqs.~(\ref{eq:dndwlx})--(\ref{eq:dndwso}),
we evaluate the observed spectrum and polarization for the expanding fireball. 

\begin{figure}[tb]
\centering
\includegraphics[width=0.48\textwidth]{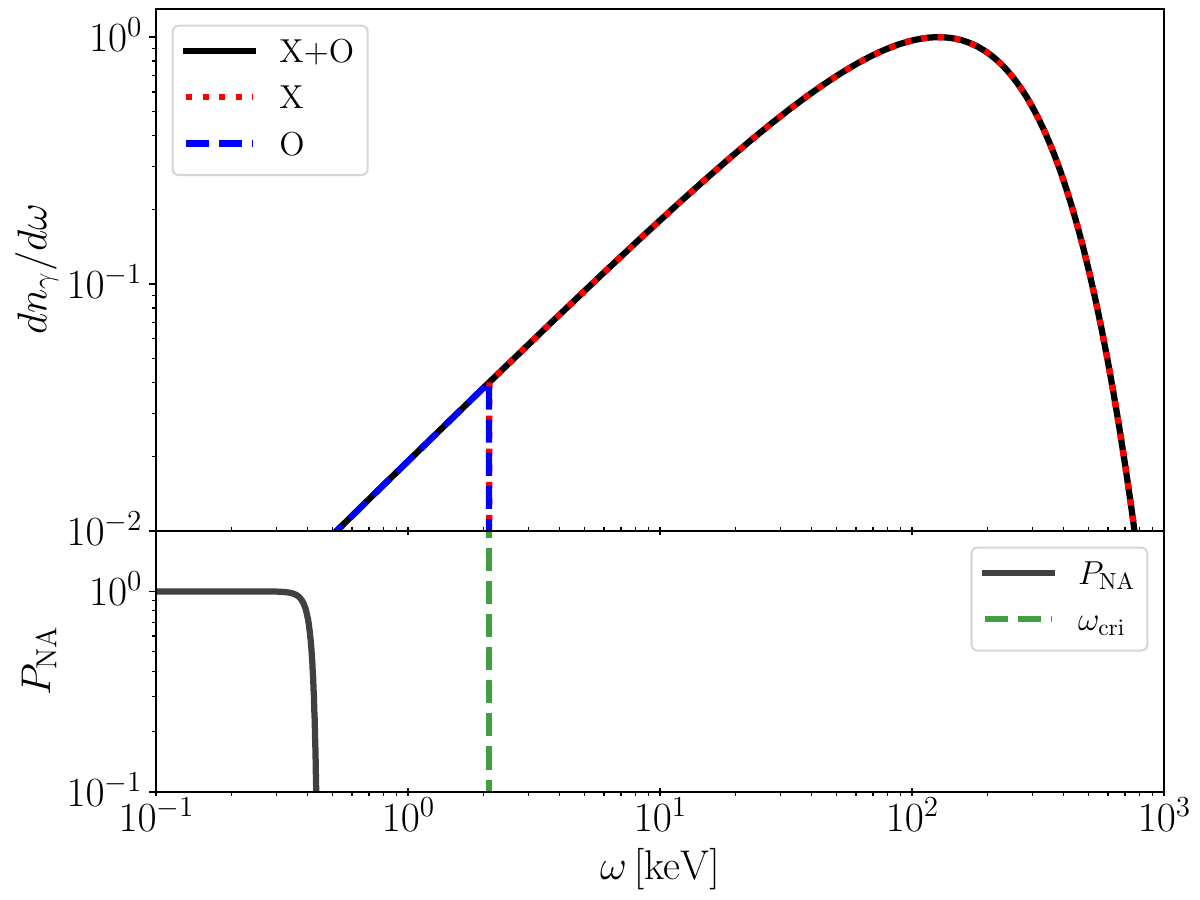}
\caption{\RaggedRight
Observed spectrum from an expanding fireball. Each line has the same meaning as the top panel of Fig~\ref{fig:trapped}.}
\label{fig:expand}
\end{figure}

%
Figure~\ref{fig:expand} shows the observed spectrum and polarization. The adopted parameters 
are $B_0 = 2\times10^{14}\,{\rm G}$, $T_0 = 80\,{\rm keV}$, $l_0 = 5\times 10^3\,{\rm cm}$, 
${\mathcal F}_{\rm exp} = 10^{-2}$ for the fireball based on \cite{WadIok2023},
and $H_{\rm res} = 10^6\,{\rm cm}$,
$\theta_{kB} = 0.1$, for the propagation.
As in the trapped-fireball case, photons with 
$\omega_{\rm ad}\lesssim \omega < \omega_{\rm cri}$ undergo an MSW-like mode conversion, in 
which the emitted X-mode photons are converted into observed O-mode photons while remaining
in the eigenmode of the propagation, $-$ mode (see Fig.~\ref{fig:K}).

\begin{figure}[tb]
    \includegraphics[width=\linewidth]{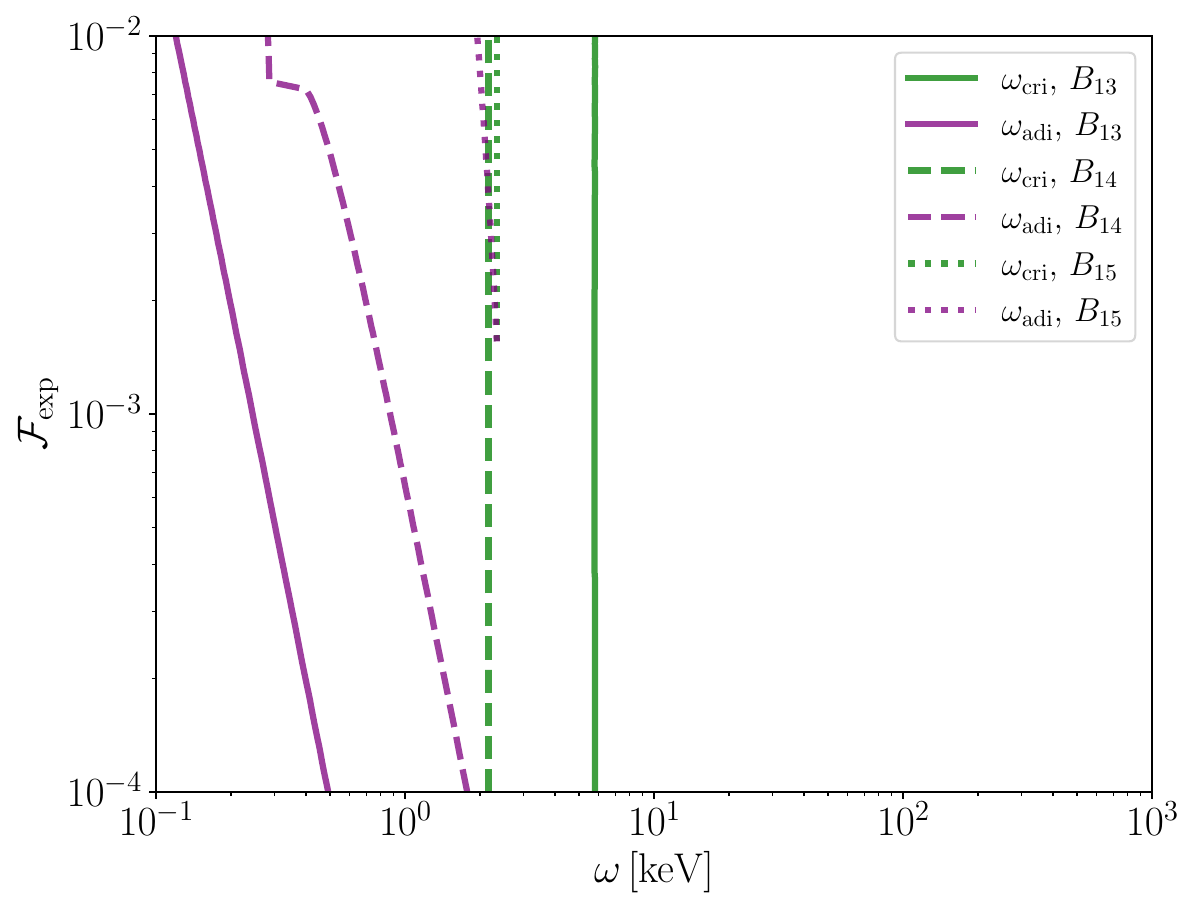}
\caption{\RaggedRight
$\omega_{\rm cri}$ (green line) and $\omega_{\rm adi}$ (purple line) for different baryon loading, ${\mathcal F}_{\rm exp}$.
Each line has the same meaning as Fig~\ref{fig:trap_para}.
}
\label{fig:exp_para}
\end{figure}

Figure~\ref{fig:exp_para} shows ${\mathcal F}_{\rm exp}$-dependences of $\omega_{\rm cri}$ 
(green line) and $\omega_{\rm adi}$ (purple line). Adopted parameters except 
${\mathcal F}_{\rm exp} $ are the same as Fig.~\ref{fig:expand}. ${\mathcal F}_{\rm exp}$ 
are chosen from $10^{-4}$ to $10^{-2}$, above which the approximation of radiation-dominant
fireball is broken (\cite{WadIok2023} for details). 
The bending around ${\mathcal F}_{\rm exp}\sim8\times 10^{-3}$ in the case of $B=10^{14}\,{\rm G}$
arises because the ion cyclotron resonance, proportional to $(1-u_b)^{2}$, affects the determination of 
$\omega_{\rm adi}$ in solving Eq.~(\ref{eq:omegacri}).
For $B=10^{15}\,{\rm G}$, $\omega_{\rm adi}$ shows little dependence on 
${\mathcal F}_{\rm exp}$ because photons escape from the hot inner region dominated by pair plasma,
due to the strong suppression on the cross section (see Eq.~\ref{eq:tau_exp}).
For the cases of $B=10^{13}\,{\rm G}$ and $10^{14}\,{\rm G}$, the MSW-like conversion into
O-mode occurs in relatively large parameter ranges compared to 
the trapped fireball, and these ranges depend on ${\mathcal F}_{\rm exp}$.
This suggests the possibility of measuring baryon-loading using 
X-ray polarization for given $B_0$.

\section{Summary and Discussion\label{sec:sum}}
%
We have developed a framework to describe vacuum resonance in the three-component plasma 
produced in baryon-loaded magnetar bursts, 
and revealed the polarization of X-ray bursts expected in future observations.
By parametrizing the plasma 
composition, we derived the normal modes, 
their polarization properties, and their dependence on composition
(Sec.~\ref{sec:modes}). 
We showed that the vacuum resonance, previously
studied in the context of magnetar persistent surface emission \cite{LaiHo2003_PRL},
can also occur in magnetar bursts. The resonance condition, given in 
Eq.~(\ref{eq:nres}), depends only on the pair number density $n_e$.
We also investigate the MSW-like mode conversion and 
the nonadiabatic transitions of normal modes at the resonance, whose 
probability, especially $f$-dependence, is given in Eqs.~(\ref{eq:PNA}) and (\ref{eq:gammares}) (Sec.~\ref{sec:conv}).
Applying the framework to analytic fireball models,
we demonstrated that vacuum resonance appears
in the observed X-ray polarization (Sec.~\ref{sec:obs}).
Throughout these works, we also clarified the connection 
between the polarization modes 
in conventional purely pair fireballs and those in baryon-loaded fireballs.
The nonadiabatic mode transition plays an important role in the X/O-mode
propagation in a pair plasma, as shown in Eq.~(\ref{eq:PNA1}).
Our results suggest that polarization observations of magnetar bursts can
provide unique diagnostics of the fireball structures and their composition.

%
In the modeling presented in Sec.~\ref{sec:obs}, the observerd
spectrum has been calculated using simplified models, with
fixed parameters, $B$, $\theta_{kB}$, ${\mathcal F}_{\rm trap/exp}$, 
and $H_{\rm res}$. Moreover, the time variability of the amount of baryon
is ignored.
In reality, the spectrum should be a superposition
of emissions with different parameter values 
and depend on time. A realistic calculation 
therefore requires time-dependent three-dimensional simulations 
that incorporate the full structure of fireball \cite{YanZha2015}.
The spectra in this work can serve as initial conditions for 
such global simulations. The observed spectrum is also expected to 
be modified by the absorption and scattering of O-mode photons, 
as well as by resonant Compton scattering in the magnetosphere
\cite{YamLyu2020,WadIok2023,ThoLyu2002}. For magnetar persistent surface emission,
rather than bursts, the presence of a hot coronal region 
composed of pair plasmas near the surface might necessitate considering 
a three-component plasma, studied in this work. 
These realistic global calculations are left for future study.

%
The polarization signature depends on the baryon content
of fireballs (Sec.~\ref{sec:obs}), so observations of X-ray
polarization may serve as a probe of the baryon loading in
fireballs. 
\cite{WadIok2023} investigated
the relation between FRB~20200428A and simultaneously observed SB within
the expanding-fireball model. They showed that mildly baryon-loaded 
fireballs produce higher kinetic energy in the plasma outflow,
which may lead to FRB emission.
On the other hand, \cite{ZhoLi2024} proposed that 
X-ray polarization of SBs can be useful to distinguish between 
expanding and trapped fireball configurations.
The polarization induced by the vacuum resonance studied in 
this paper may provide a means to constrain the 
the amount of baryon lading on fireballs (see Fig.~\ref{fig:exp_para})
by the observation of X-ray polarization.
By combining this work and \cite{ZhoLi2024},
future simultaneous observation of 
FRBs and SB polarizations would enable the determination of the fireball
configuration as well as its baryon loading, providing a powerful
probe of the still-unclear FRB emission mechanism.

\begin{acknowledgments}
We thank
K. Asano, T. Kawashima, and  J. Shimoda 
for comments leading this research area
and occasional discussions. 
We also thank the anonymous referee for many valuable comments that helped improve the manuscript.
This work is supported by JSPS KAKENHI Grant-in-Aid for Scientific Research No.~JP25KJ0024 and No.~JP25K17378.
\end{acknowledgments}
%
%
\bibliography{cite}
\end{document}